\documentclass[
10pt,aps,
notitlepage, 
bibnotes,
prl,
twocolumn,
showpacs,preprintnumbers,superscriptaddress,
amsmath,amssymb,floatfix]{revtex4-2}

\usepackage{natbib}  

\usepackage{graphicx}
\usepackage{dsfont}
\usepackage{dcolumn}
\usepackage{bm}
\usepackage{color}
\usepackage{xcolor}
\usepackage{url}
\usepackage{tabularx}
\usepackage{array}
\usepackage{booktabs}
\usepackage{comment}
\usepackage{hyperref}
\usepackage{footnote}
\usepackage{lipsum}
\usepackage[normalem]{ulem}
\usepackage{booktabs}
\usepackage{braket}

\DeclareSymbolFont{md}{OMX}{mdput}{m}{n} 
\DeclareMathSymbol{\intop}{\mathop}{md}{90}

\definecolor{colorA}{rgb}{0, 0, 1}

\makeatletter
  \def\my@tag@font{\normalsize}
  \def\maketag@@@#1{\hbox{\m@th\normalfont\my@tag@font#1}}
  \let\amsmath@eqref\eqref
  \renewcommand\eqref[1]{{\let\my@tag@font\relax\amsmath@eqref{#1}}}
\makeatother

\begin{document}

\title{Co-existing magnetization reversal mechanisms in shakti spin ice systems}

\author{Vladyslav M. Kuchkin}
\email{vladyslav.kuchkin@uni.lu}
\affiliation{Department of Physics and Materials Science, University of Luxembourg, L-1511 Luxembourg, Luxembourg}

\author{Unnar~B.~Arnalds}
\affiliation{Science Institute and Faculty of Physical Sciences, University of Iceland, 107 Reykjav\'ik, Iceland}

\author{Hannes J\'onsson}
\affiliation{Science Institute and Faculty of Physical Sciences, University of Iceland, 107 Reykjav\'ik, Iceland}

\author{Pavel~F.~Bessarab}
\email{pavel.bessarab@lnu.se}
\affiliation{Science Institute and Faculty of Physical Sciences, University of Iceland, 107 Reykjav\'ik, Iceland}
\affiliation{Department of Physics and Electrical Engineering, Linnaeus University, SE-39231 Kalmar, Sweden}

\date{\today}

\begin{abstract}
The switching mechanisms in artificial spin ice systems are investigated with focus on shakti and modified shakti lattices. 
Minimum energy paths are calculated using the geodesic nudged elastic band (GNEB) method implemented with a micromagnetic description of the system, including the internal magnetic structure of the islands
and edge modulations.
Two switching mechanisms, uniform magnetization rotation and domain wall formation, are found to have comparable activation energy. 
The preference for one over the other 
depends strongly on the saturation magnetization and the magnetic ordering of neighboring islands. 
Surprisingly, these mechanisms can coexist, leading to an enhanced probability of magnetization reversal.
These results provide valuable insight that can help control
internal magnetization switching processes in spin ice systems and help predict their thermodynamic properties.

\end{abstract}
\maketitle

\textit{Introduction.} 
Artificial spin ice (ASI) systems~\cite{Wang2006,Heyderman2013, Gilbert2016,Nisoli2017, Rodriguez-Gallo_NatComm_2023} represent interacting nanomagnets demonstrating complex emergent behaviors due to their frustrated lattice structure absent in ordinary magnetic materials.
Due to their unique magnetic properties, spin ice systems have potential for various applications in next-generation data storage devices\cite{schiffer_APL_2021}, neuromorphic computations\cite{Gartside_NatNano_2022,Hu_NatComm_2023}, and quantum computing~\cite{King2021}.
Magnetic ordering and dynamics observed in artificial spin ice systems have been a rich playground for investigations of a multitude of phenomena in interacting nanomagnetic systems of different geometries \cite{skjaervo_NRP_2020, Ostman_NP_2017, Morrison_NJP_2013,arnalds_NJP_2016}. Such systems have been used for investigating ordering phenomena such as frustration~\cite{Morrison2013,Chern2013,Gilbert2015,Drisko2017} and higher-order vertex interactions and how they are modified by the geometry of the islands as well as spin wave dynamics both internally in the islands and between islands \cite{Skovdal_PRB_2023, PhysRevB.100.214425, gartside2021reconfigurable}. 

\begin{figure*}
    \centering
    \includegraphics[width=17.8cm]{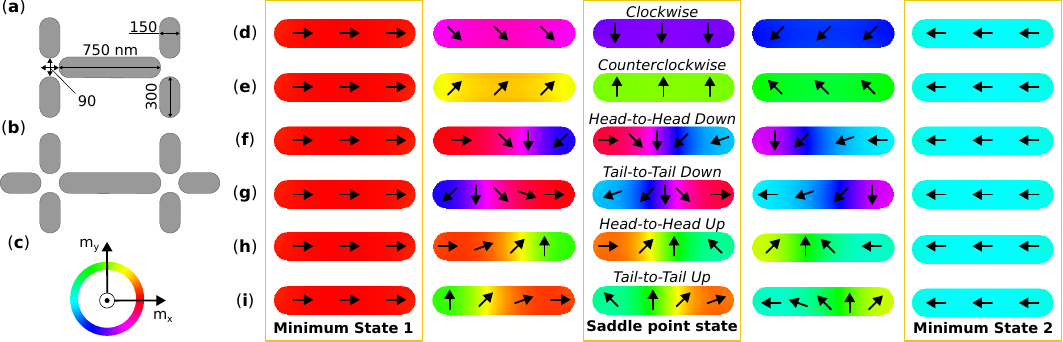}
    \caption{\textbf{Island magnetization reversal mechanisms}.
   (\textbf{a}) and (\textbf{b}) show the geometry of the shakti and modified shakti arrangements used for the calculations of the reversal barrier for elongated islands. 
    The color code used to indicate the magnetization direction is illustrated in (\textbf{c}).
    (\textbf{d})-(\textbf{i}) show the six possible switching mechanisms of the elongated island: uniform rotation (clockwise and counterclockwise) and domain wall formation with head-to-head and tail-to-tail types, with down ($-\bm{e}_\mathrm{y}$) and up ($+\bm{e}_\mathrm{y}$) directions of the central spin.
    }
    \label{Fig1}
\end{figure*}

Several articles have dealt with the thermal dynamics involved in artificial spin ice systems, e.g., as the systems transition from a frozen ordered state to a thermally active state or can be thermally activated by heating to an elevated temperature \cite{Arnalds_APL_2012, Vassilios_NN_2014}.
These dynamics arise from thermally activated reversals of the magnetization of individual islands from the two opposing energy-minimum states.
These studies highlight the importance of the energy barrier on the ordering in the system as any potential dynamics in these systems are directly coupled to the reversal mechanism and energy barrier of individual islands. 
This important role is e.g., exemplified in shakti [Fig.~\ref{Fig1}(\textbf{a}), (\textbf{b})] and Saint George spin ice lattices composed of islands with different reversal energy barriers which freeze into a low-temperature state at different temperatures governed by their reversal barrier \cite{Stopfel_PRB_2018, Stopfel_PRM_2021, Gilbert_NP_2014}. 

In many cases, simple models can be employed to determine the energy barrier for an island reversal, such as determining them directly using the Stoner-Wolfarth model for ellipsoidal islands \cite{Osborn_PR_1945} as well as the micromagnetic modeling of the energy barrier assuming uniform rotation of magnetization. These model calculations are limited as they do not capture thermal effects and exclude non-uniform magnetization reversal mechanisms from consideration.
Such calculations may also neglect any potential divergences occurring at the edges of the islands arising from interactions with neighboring islands at the vertex and how they potentially alter the magnetization switching between the two ground states. In addition to the reversal barrier inherent in individual islands, the vertex arrangements of islands have recently been shown to comprise an additional modification of the energy landscape within individual islands through divergences at the end of the islands, rich in dynamics and adding to the ordering phenomena of ASI systems \cite{Skovdal_PRB_2023, Sloetjes_APL_2021, Sloetjes_PRB_2022}. 

In this study, we develop and apply micromagnetic-based minimum energy path (MEP) calculations to determine energy barriers in square and shakti artificial spin ice. Using our model, we obtain full energy barrier profiles and intermediate states of the systems as they traverse between energy-minimum states. Determining the reversal energy barrier for individual islands as a function of island magnetization, we observe a crossover between the reversal mechanisms from a uniform rotation to a domain wall formation. 
Above this bifurcation point, our modeling shows both the reversal mechanisms to be stabilized, enabling the reversal to occur through two pathways, potentially affecting the dynamics of the system, especially close to the bifurcation point. Altering the magnetization of neighboring islands we observe the reversal mechanism to be strongly affected by the magnetization direction of neighboring islands both for individual islands as well as for smaller-scale energy barriers between combined island divergence states at vertices of four islands. 

\textit{Model and method.} The Hamiltonian used for describing spin ice systems is of the following form,
\begin{equation}
\!\!\mathcal{H}=l\int\left(\mathcal{A}\left[\left(\partial_\mathrm{x}\bm{m}\right)^{2}+\left(\partial_\mathrm{y}\bm{m}\right)^{2}\right]-\dfrac{1}{2}M_\mathrm{s}\bm{b}_\mathrm{d}\cdot\bm{m}\right)\mathrm{d}S,\label{Hamiltonian}
\end{equation}
where $\bm{m}$ is the unit vector in the direction of local magnetization, $\mathcal{A}$ is the exchange stiffness constant, $M_\mathrm{s}$ is the saturation magnetization, $\bm{b}_\mathrm{d}$ is the demagnetizing field and $l$ is the thickness of the islands. 
We fix $l=3.3$~nm and, thus, can neglect magnetization variation along the island thickness, i.e., $\bm{m}=\bm{m}(x,y)$.
The exchange constant equals $13$~pJ/m, which is a standard value for permalloy~\cite{Vansteenkiste2014}. 
The saturation magnetization value we vary in the range $[100,350]$~kA/m in most simulations, and only for calculating energy barriers and reversals for edge modulation states, we consider $M_\mathrm{s}=800$~kA/m.
The geometry of spin ice systems is shown in Fig.~\ref{Fig1} for shakti (\textbf{a}) and modified shakti (\textbf{b}) islands.
For both cases, we focus our study on the switching of the elongated island while the smaller islands remain in their primary magnetization.
We use a standard color code [Fig.~\ref{Fig1} (\textbf{c})] to visualize the magnetization within the islands.

To obtain equilibrium states in the spin ice systems, we minimize the energy~Eq.~\eqref{Hamiltonian} in the Mumax3 software~\cite{Mumax} relying on the steepest descent methods, i.e. utilizing the effective field $\bm{B}_\mathrm{eff} = -M_\mathrm{s}^{-1}\delta\mathcal{H}/\delta\bm{m}$.
Calculation of MEPs and energy barriers between any two stable states is done with the geodesic nudged elastic band (GNEB) method~\cite{Bessarab2015, Bessarab2017}, which is implemented in Mumax3 and available in a GitHub repository~\cite{gneb_git}.
Recently performed GNEB calculations for complex magnetic spin textures such as skyrmions~\cite{Kuchkin2022,Kuchkin2022_2}, hopfions~\cite{Kuchkin2023,Sallermann2023}, and magnetic bubbles~\cite{Savchenko2023} in multilayer systems demonstrate its high efficiency in studying a wide variety of textures.  
According to the method, we consider a set of $N$ states, referred to as images, $\bm{M}_{\nu}$, where the initial $\nu=1$ and final $\nu=N$ images correspond to equilibrium points, while all intermediate $1<\nu<N$ images are movable points.
Each of the movable images, $\bm{M}_\nu$, is subject to 
a force $\bm{F}_{\nu}$ arising from the effective field $\bm{B}_\mathrm{eff}$ and from adjacent images $\bm{M}_{\nu-1}$ and $\bm{M}_{\nu+1}$. 
The latter force is a spring force and is proportional to the geodesic distance between images.
Applying the velocity projection optimization method, the forces $\bm{F}_{\nu}$
are zeroed to a given tolerance.

\begin{figure*}
    \centering
    \includegraphics[width=17.5cm]{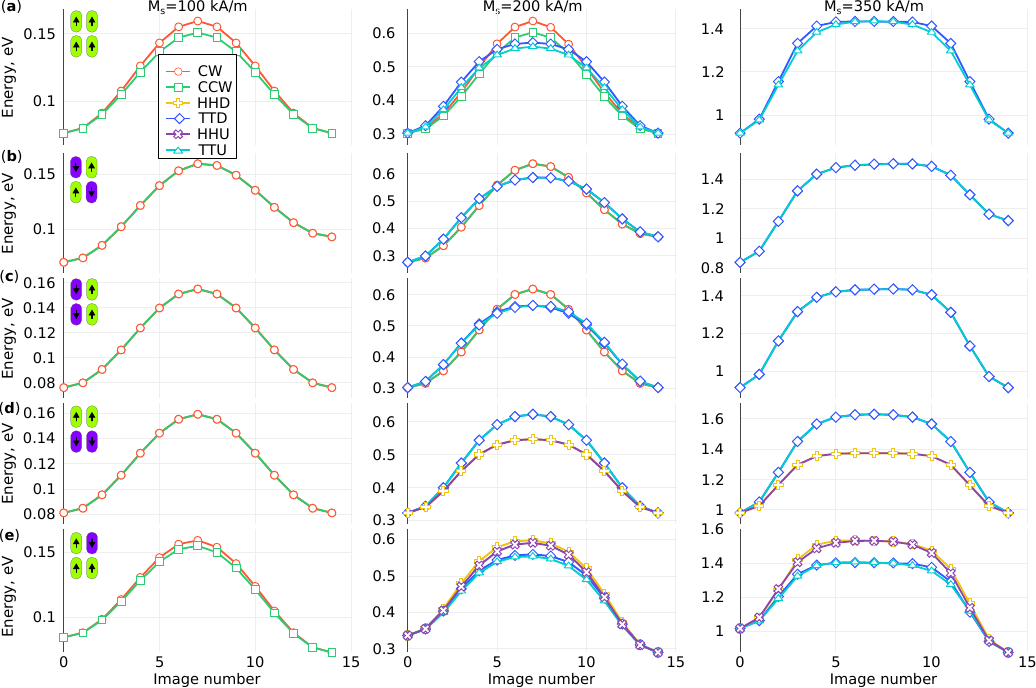}
    \caption{\textbf{Switching mechanisms in the shakti system}. 
    (\textbf{a})-(\textbf{e}) show MEPs for five cases for different magnetizations of the small islands and values of $M_\mathrm{s}=100, 200$ and $350$~kA/m. 
    The orientation of the four small islands, surrounding the elongated island, is given in the top-left insets.
    CW and CCW switchings are present only for $M_\mathrm{s}=100$~kA/m (\textbf{a})-\textbf{(e)} and  for $M_\mathrm{s}=200$~kA/m (\textbf{a})-\textbf{(c)}.
    HHD, TTD, HHU, and TTU switchings are present for $M_\mathrm{s}=200$~kA/m and $M_\mathrm{s}=350$~kA/m.
    In cases (\textbf{a})-(\textbf{c}), MEPs for HHD (HHU) coincide with those for TTD (TTU), while for geometries (\textbf{d}), (\textbf{e}), MEPs for HHD (TTD) are comparable to HHU (TTU). 
    }
    \label{Fig2}
\end{figure*}

The GNEB method makes it possible to identify the
saddle point configuration, as well as to determine the whole MEP, which shows the mechanism of the transition.
The maximum of the energy along the path minus the energy of the initial state gives the energy barrier for the transition and the transition rate can be calculated as a function of temperature using the harmonic approximation to the transition state theory~\cite{Bessarab2012, Bessarab2013}.
The estimated rates of the elementary transitions can be used to calculate the lifetime of the magnetic states and switching time.

\textit{Saddle points.} Every island in the spin ice system is magnetized along the primary axis due to the shape anisotropy. 
As shown in Fig.~\ref{Fig1}, switching between right- and left-magnetized islands, denoted as Minimum States 1 and 2, can occur via uniform rotation (\textbf{d}), (\textbf{e}) or via domain wall formation (\textbf{f)}-(\textbf{i}) represented by the saddle point configurations and illustrated in the figure. 
Here, we ignore multi-domain switching mechanisms~\cite{ChavesOFlynn2009} as they have higher energy barriers in the range of saturation magnetization values we consider and are only relevant in the case of high $M_\mathrm{s}$. 
In the case of planar magnets, the characteristic size of the domain wall is given by the exchange length, $l_\mathrm{ex}=\sqrt{2\mathcal{A}/(\mu_{0}M_\mathrm{s}^2)}$, which can be used as a good estimate for comparing the domain wall width to the size of the islands.

For a single island switching, clockwise (CW) and counterclockwise (CCW) uniform rotations (\textbf{d}), (\textbf{e}) have identical energy barriers due to the symmetry of the system.
The same holds for four types of switching via the domain wall formation (\textbf{f})-(\textbf{i}).
However, in spin ice systems, the way neighboring islands are magnetized will differentiate these switching cases.
Due to the presence of the small islands, energy barriers for CW and CCW switchings do not generally coincide.
The same holds for the domain wall formation mechanisms.
We distinguish four types of domain wall transitions: head-to-head down (HHD), tail-to-tail down (TTD), head-to-head up (HHU), tail-to-tail up (TTU) shown in Fig.~\ref{Fig1} (\textbf{f})-(\textbf{i}).

\begin{figure}
    \centering
    \includegraphics[width=8.5cm]{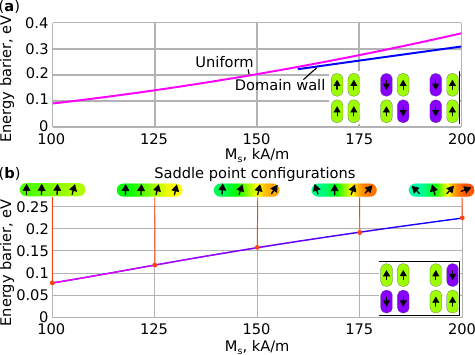}
    \caption{\textbf{Primary mechanism for magnetization reversal}. (\textbf{a}) and (\textbf{b}) show the calculated energy barriers for the magnetization reversal of the elongated islands as a function of saturation magnetization.
    (\textbf{a}) and (\textbf{b}) correspond to different magnetization directions of the four neighboring small islands, as provided in the bottom right insets.
    In the cases shown in (\textbf{a}), there are two mechanisms -- uniform rotation and switching via the domain wall formation -- that have a crossing point at $M_\mathrm{s}\simeq 160$~ kA/m.
    In the cases shown in (\textbf{b}), there is only one primary switching mechanism representing a mixture of those two.
    The inset images above (\textbf{b}) show saddle point states for selected values of $M_\mathrm{s}$.
    }
    \label{Fig3}
\end{figure}

\textit{Shakti geometry.}
Considering the symmetry of the shakti geometry, there are five unique cases for the magnetic arrangement of the surrounding small islands when considering the energy and reversal of the elongated island. An overview of energetically representative states is shown in Fig.~\ref{Fig2}~(\textbf{a})-(\textbf{e}).
Each of those energetically identical cases has its own symmetry, which is reflected in the switching mechanisms and non-equivalence of the MEPs.
The general feature of all MEPs is that uniform rotation switching (CW and CCW) appears only at low saturation magnetization values, while switching via domain wall formation (HHD, TTD, HHU, and TTU) is present at higher values of $M_\mathrm{s}$.
However, at some intermediate values of $M_\mathrm{s}$, all types of reversal mechanisms can co-exist depending on the magnetization of the neighboring islands and their symmetry. 
Note that in a generic model of a bistable magnet characterized by exchange and anisotropy, uniform rotation of magnetization cannot coexist with other reversal modes (see Supplemental Material~\cite{Suppl}). Therefore, the coexistence of switching mechanisms arises from magnetostatic interactions beyond shape anisotropy, as well as the influence of neighboring islands.

The configurations shown in Fig.~\ref{Fig2}(\textbf{a}) and Fig.~\ref{Fig2}(\textbf{e}) are characterized by the lowest symmetry.
This results in that CW and CCW switchings have different energy barriers, or in other words, their MEPs are not equivalent.
Different energy barriers are also observed for switching via domain wall formation. In the case shown in Fig.~\ref{Fig2}(\textbf{a}), energy profiles along MEPs for HHD (HHU) and TTD (TTU) coincide, but a minor splitting is observed for HHD and HHU, while for the case shown in Fig.~\ref{Fig2}(\textbf{e}), all  MEPs correspond to different energy barriers. 
Configurations shown in Fig.~\ref{Fig2}(\textbf{b})-(\textbf{d}) are more symmetric and MEPs for CW and CCW transitions are equivalent.
For cases shown in Fig.~\ref{Fig2}(\textbf{b}) and Fig.~\ref{Fig2}(\textbf{c}), all switching mechanisms involving domain wall formation are equivalent. 
However, the head-to-head and tail-to-tail domain wall switching mechanisms are not equivalent for the case shown in  Fig.~\ref{Fig2}(\textbf{d}). 
Additionally, at $M_\mathrm{s}=200$~kA/m, CW and CCW transitions are present only for geometries (\textbf{a})-(\textbf{c}).
Reversal mechanism simulations for the modified shakti geometry revealed similar results to those for the shakti geometry, with both uniform rotation and domain wall reversal mechanisms present.
The resulting energy barriers for varying magnetization are provided in the Supplemental Material~\cite{Suppl}. 

In the theory of magnetization reversal, transitions with the lowest energy barrier are typically of primary interest.
By comparing the cases shown in Fig.\ref{Fig2}, it becomes evident that uniform rotation and switching via domain wall can both represent the primary switching mechanism at different $M_\mathrm{s}$.
Consequently, the coexistence of different switching mechanisms can be anticipated at a specific magnetization saturation. In the following, we explore this in more detail.

\textit{Primary switching mechanisms.}
As illustrated in the discussion above, the GNEB method reveals different switching mechanisms for the elongated island in a shakti geometry depending on the saturation magnetization value and the arrangement of neighboring islands. 
To investigate the reversal mechanism and its evolution systematically, we performed multiple runs of GNEB simulations for different values of $M_\mathrm{s}$ and compared energy barriers for uniform reversal switching and switching via domain wall reversal. 
Graphs summarizing the results are shown in Fig.~\ref{Fig3}.
For states shown in Fig.~\ref{Fig3}(\textbf{a}), uniform rotation is the only mechanism for $M_\mathrm{s}<160$ kA/m, and switching through the domain wall formation is not present. 
For $M_\mathrm{s}>160$~kA/m, the simulations revealed two mechanisms to be stable: uniform rotation and switching through domain wall formation with a lower energy.
For the states shown in Fig.~\ref{Fig3}(\textbf{b}), there is no such strict distinction, and for the intermediate value of $ M_\mathrm{s}$, the saddle point rather represents a combination of a uniform rotation state and domain wall reversal.
The corresponding saddle point configurations are provided in Fig.~\ref{Fig3}(\textbf{b}).

\begin{figure}
    \centering
    \includegraphics[width=8.5cm]{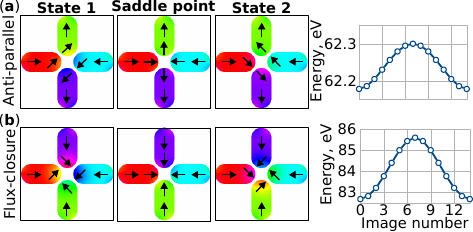}
    \caption{\textbf{Edge modulations}. (\textbf{a}) and (\textbf{b}) contain spin textures corresponding to the anti-parallel and flux-closure edge modulations, respectively.
    The MEPs calculated for $M_\mathrm{s}=800~\mathrm{kA/m}$. 
    The anti-parallel states are separated by an energy barrier of $\Delta E=0.1$~eV, while the flux-closure edge modulation states are separated by a barrier of $\Delta E = 2.9$~eV.
}
    \label{Fig4}
\end{figure}

\textit{Edge modulations.}
To better understand the effect of neighboring islands on the reversal mechanism, we investigated edge modulations in so-called Type I and Type IV vertex states in artificial square spin ice vertices shown in Fig.~\ref{Fig4}.
These states have been shown to have anti-parallel (\textbf{a}) and flux-closure (\textbf{b}) edge modulations.\cite{Skovdal_PRB_2023, Sloetjes_APL_2021, Sloetjes_PRB_2022}
In each case, the symmetric edge modulation states are separated by energy barriers, which can be identified by the GNEB method.
As we can deduce from Fig.~\ref{Fig4} (\textbf{c}), these barriers are substantially smaller in magnitude than those discussed previously, and in order to observe them, higher values of saturation magnetization are needed, so we utilize $M_\mathrm{s}=800$~kA/m for simulations.
Thus, we do not expect a significant influence of such edge modulations on the energy barrier magnitude for $M_\mathrm{s}\in[100, 350]$~kA/m, as discussed above.
However, these edge modulations will affect the reversal pathway, i.e., whether the reversal occurs through a CW or CCW reversal for uniform rotation or via HHU, HHD, TTU, or TTD reversal for domain wall formation.
At higher values of $M_\mathrm{s}$, when the edge modulations can be observed, one has to take into account more diverse possibilities for saddle point configurations, e.g., states with multiple domain walls, vortices, etc., especially for elongated islands. 
Analysis of such transitions is beyond the scope of the current work and will be the subject of further investigation.

\textit{Conclusions.}
In conclusion, we study the switching mechanisms of shakti spin ice and modified shakti geometries using the geodesic nudged elastic band method implemented in combination with a micromagnetic description of the system.
Our findings reveal two reversal mechanisms -- uniform rotation and switching via domain wall formation. The corresponding energy barriers depend on the material's saturation magnetization and on the magnetization arrangement of neighboring islands.
Critical values for $M_\mathrm{s}$
have been identified
at which a change in the primary reversal mechanism occurs, resulting in a bifurcation point of co-existing reversal mechanisms.
Finally, the role of low-energy barriers corresponding to edge modulations in the modification of these primary mechanisms
is revealed.
The reversal pathways are found to be modified by the magnetic arrangement of neighboring islands, and this is particularly important to take into account for materials with high $M_\mathrm{s}$. 
In such cases, there can even be the 
possibility of switching mechanisms via multi-domain wall formation.

%
\begin{acknowledgments} V.M.K. acknowledges financial support from the National Research Fund of Luxembourg under the CORE Grant No.~C22/MS/17415246/DeQuSky. This work was supported by funding for the project Magnetic Metamaterials from the Icelandic Research Fund grant no. 2410333,
by the University of Iceland Research Fund, by the Swedish Research Council (grant No. 2020-05110), and by the Crafoord Foundation (grant No. 20231063).
The authors acknowledge Hendrik Schrautzer, Moritz Sallermann, Jan Masell, and Jonathan Leliaert for valuable discussions.
\end{acknowledgments}

\bibliographystyle{apsrev4-2}
\bibliography{main_arxiv}

\newpage

\onecolumngrid
\newpage
\begin{center}
   \textbf{Supplemental Material for ``Co-existing magnetization reversal mechanisms in shakti spin ice systems''}\newline 
\end{center}

\section{Crossover between switching mechanisms in a uniaxial magnet}\label{AppendixB}

We consider a one-dimensional model of a ferromagnet equipped with the uniaxial anisotropy along the system's axis. The energy of the system reads:
\begin{equation}
\label{seq:ene}
    E=\int_0^L\left[A\left(\frac{d\bm{m}}{dx}\right)^2-Km_x^2\right]dx,
\end{equation}
where $L$ is the system size and $K>0$ is an easy-axis anisotropy constant. There are two stable states with $\bm{m}$ pointing either along or opposite to the $x$-axis. In the macrospin approximation, there is only one (degenerate) saddle point corresponding to $\bm{m}$ orthogonal to the $x$-axis. The task is to identify under what conditions this state is a saddle point of the functional given by Eq. (\ref{seq:ene}). Rewriting the functional in terms of the spherical coordinates $\bm{m}=(\sin\theta\cos\varphi,\sin\theta\sin\varphi,\cos\theta)$ and expanding it around the the point $\theta_0=\pi/2$, $\varphi_0=\pi/2$ to the second order of the deviation yields: 
\begin{equation}
\label{seq:approx}
    E\approx E_0+\int_0^L\bm{\xi}^T D\bm{\xi} dx,
\end{equation}
where $E_0=0$ is the energy of the collinear state with $\bm{m}(x)$ perpendicular to the $x$-axis, $\bm{\xi}^T=[\delta\theta(x),\delta\varphi(x)]$ describes the deviation, and the operator $D$ is given by the following equation:
\begin{equation}
    \label{seq:D}
    D\equiv \begin{pmatrix}
        -A\dfrac{d^2}{dx^2} & 0\\
        0 & -A\dfrac{d^2}{dx^2} - K.
    \end{pmatrix}
\end{equation}
The eigenvalues of $D$ are fixed by the boundary condition $\bm{\xi}(0)=\bm{\xi}(L)=0$:
\begin{equation}
    \varepsilon_n^\pm=\frac{A\pi^2n^2}{L^2}-\frac{1}{2}(K\pm K), 
\end{equation}
where $n$ is an integer number. 
There is at least one zero eigenvalue $\varepsilon_0^-=0$ that corresponds to the zero mode -- rotation of the magnetization around the $x$ axis. There is at least one negative eigenvalue, $\varepsilon_0^+=-K$. If this is the only negative eigenvalue, the collinear state corresponds to a (degenerate) first order saddle point and a true mechanism of magnetization reversal. This is the case when $\varepsilon_1^{(+)}=A\pi^2/L^2-K>0$, which is realized when $L<L_0$ with $L_0$ being the domain wall width: $L_0=\pi\sqrt{A/K}$. However, the second eigenvalue becomes negative when $L>L_0$. This indicates a crossover between the mechanisms of the reversal: the collinear rotation breaks down, and non-uniform rotation of magnetization is established.

\section{Switching mechanisms in modified shakti spin ice systems}\label{AppendixA}

\begin{figure*}
    \centering
    \includegraphics[width=17.4cm]{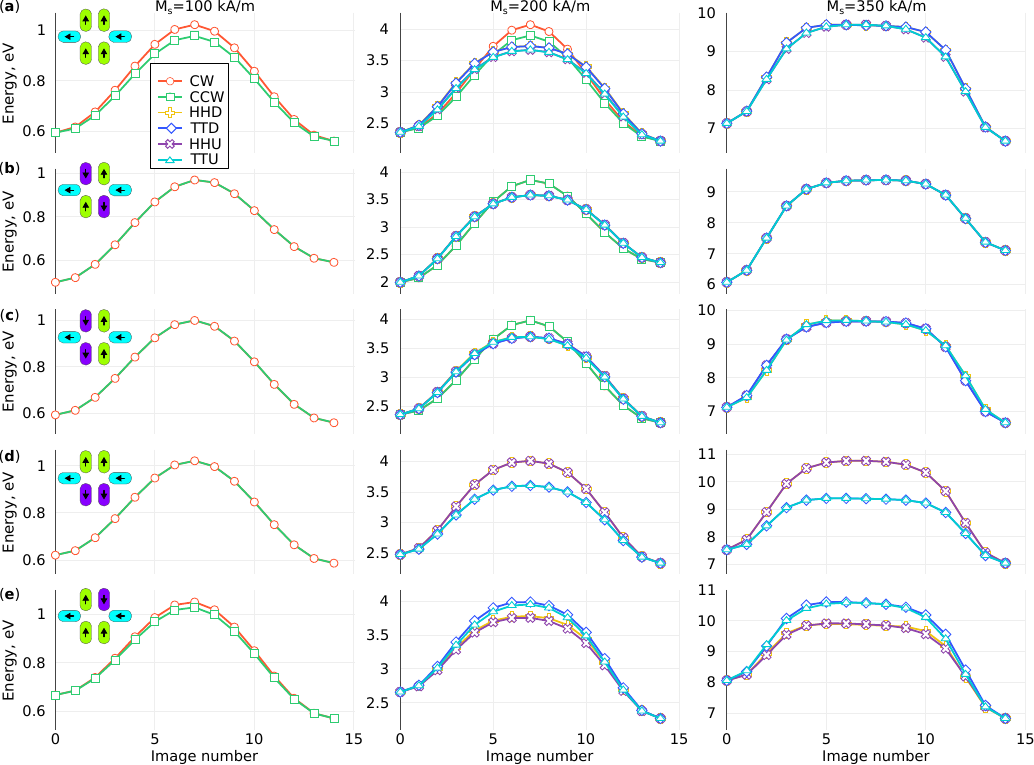}
    \caption{\textbf{Switching mechanisms in the modified shakti system: type 1}. 
    (\textbf{a})-(\textbf{e}) show MEPs for five cases for different magnetizations of the small islands and values of $M_\mathrm{s}=100, 200$ and $350$~kA/m. 
    The orientation of the six neighboring small islands is given in the top-left insets.
    CW and CCW switchings are present only for $M_\mathrm{s}=100$~kA/m (\textbf{a})-\textbf{(e)} and  for $M_\mathrm{s}=200$~kA/m (\textbf{a})-\textbf{(c)}.
    HHD, TTD, HHU, and TTU switchings are present for $M_\mathrm{s}=200$~kA/m and $M_\mathrm{s}=350$~kA/m.
    In cases (\textbf{a})-(\textbf{c}), MEPs for HHD (HHU) coincide with those for TTD (TTU), while for geometries (\textbf{d}), (\textbf{e}), MEPs for HHD (TTD) are comparable to HHU (TTU).}
    \label{fig_shatki_modif1}
\end{figure*}

\begin{figure*}
    \centering
    \includegraphics[width=17.4cm]{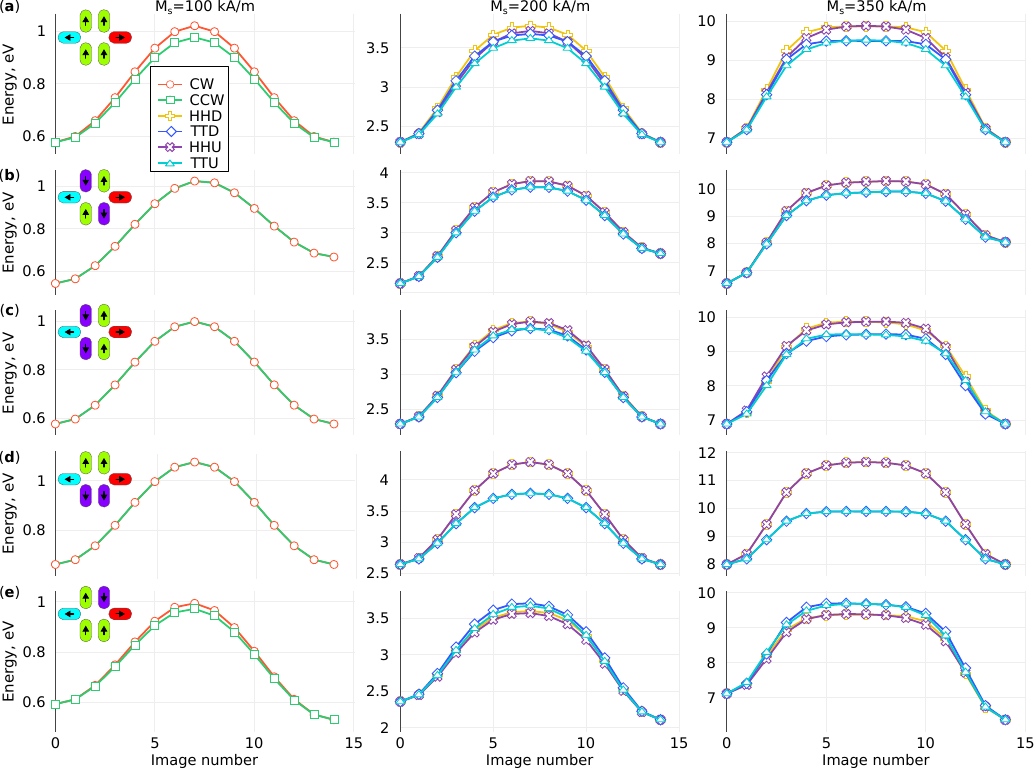}
    \caption{\textbf{Switching mechanisms in the modified shakti system: type 2}. 
    (\textbf{a})-(\textbf{e}) show MEPs for five cases for different magnetizations of the small islands and values of $M_\mathrm{s}=100, 200$ and $350$~kA/m. 
 The orientation of the six neighboring small islands is given in the top-left insets.
    CW and CCW switchings are present only for $M_\mathrm{s}=100$~kA/m.
    HHD, TTD, HHU, and TTU switchings are present for $M_\mathrm{s}=200$~kA/m and $M_\mathrm{s}=350$~kA/m.
    In cases (\textbf{a})-(\textbf{c}), MEPs for HHD (HHU) coincide with those for TTD (TTU), while for geometries (\textbf{d}), (\textbf{e}), MEPs for HHD (TTD) are comparable to HHU (TTU).}
    \label{fig_shatki_modif2}
\end{figure*}

Performing GNEB simulations for the modified shakti geometry, we obtain magnetization reversal mechanisms similar to the case of the shakti geometry.
In addition to the five orientations of the small vertical islands, we must consider two types of lateral island magnetizations. 
We refer to cases in which the lateral islands are identically or oppositely magnetized as Type 1 and Type 2, respectively. 
The results of simulations of the Type 1 modified shakti geometry are shown in Fig.~\ref{fig_shatki_modif1} and qualitatively are similar to the case of the shakti geometry discussed in the main text.
The main difference we can point out here is the energy difference between the initial and final states, which arise from the interaction of the long island to the additional lateral islands.
The results of simulations of the Type 2 modified shakti geometry are shown in Fig.~\ref{fig_shatki_modif2}.
The significant difference of this case compared to the shakti geometry is the absence of the uniform rotation switching at $M_\mathrm{s}=200$~kA/m for all types of island orientations (\textbf{a})-\textbf{(e)}.
This indicates that the primary switching mechanism changes at lower values of $M_\mathrm{s}$ compared to results shown in Fig.~\ref{Fig3} in the main text.

\end{document}